%% file: main.tex
\documentclass[sigplan,10pt]{acmart}

\acmConference[]{}{}{}
\acmYear{}
\acmISBN{} 
\acmDOI{} 
\startPage{1}

\setcopyright{none}

\usepackage{booktabs}   
\usepackage{subcaption} 
\usepackage{program}
\input{macros}

\input{commands}

\begin{document}

\title{\tech{}: Combining Numerical Optimization with SAT Solving}         


\author{Jeevana Priya Inala}
\affiliation{
  \institution{MIT}            
}
\email{jinala@csail.mit.edu}          

\author{Sicun Gao}
\affiliation{
  \institution{UCSD}           
}
\email{sicung@ucsd.edu}         

\author{Soonho Kong}
\affiliation{
  \institution{Toyota Research Institute}           
}
\email{soonho.kong@tri.global}         

\author{Armando Solar-Lezama}
\affiliation{
  \institution{MIT}           
}
\email{asolar@csail.mit.edu}         

\renewcommand{\shortauthors}{J.P. Inala, S. Gao, S. Kong, and A. Solar-Lezama}

\begin{abstract}
\input{abstract}
\end{abstract}

\ccsdesc[500]{Software and its engineering~General programming languages}
\ccsdesc[300]{Social and professional topics~History of programming languages}


\maketitle

\input{intro}

\input{overview}
\input{synthesis}

\input{solver }

\input{eval}
\input{related}


\bibliography{bibliography}

\end{document}

%% file: macros.tex
\usepackage{courier}            
\usepackage[scaled]{helvet} 
\usepackage{xspace}
\usepackage{graphicx}
\usepackage{amsmath}
\usepackage{multirow}
\usepackage[normalem]{ulem}
\usepackage{booktabs,makecell}
\usepackage{diagbox}
\useunder{\uline}{\ul}{}

\usepackage{amsthm}
\usepackage{array}
\newcolumntype{L}[1]{>{\raggedright\let\newline\\\arraybackslash\hspace{0pt}}m{#1}}
\newcolumntype{C}[1]{>{\centering\let\newline\\\arraybackslash\hspace{0pt}}m{#1}}
\newcolumntype{R}[1]{>{\raggedleft\let\newline\\\arraybackslash\hspace{0pt}}m{#1}}
\usepackage{proof}
\usepackage{amscd}
\usepackage{amssymb}
\usepackage{float}
\usepackage{epsfig}
\usepackage{xcolor}
\usepackage{color}
\usepackage{appendix}
\usepackage{graphics}
\usepackage{wrapfig,lipsum,booktabs}

\usepackage{subcaption}
\usepackage{pgf}
\usepackage{tikz}
\usepackage{adjustbox}     
\usepackage{stmaryrd}
\usepackage[T1]{fontenc}

\usepackage[noend]{algpseudocode}

\usepackage{tikz}
\usetikzlibrary{fit,backgrounds}

\newcommand{\tech}{\textsc{ReaS}}
\newcommand{\sys}{\textsc{ReaS}}
\newcommand{\Sk}{\textsc{Sketch}}

\newcommand{\hide}[1]{} 
\renewcommand{\C}[1]{\lstinline!#1!}
\renewcommand\t[1]{\text{#1}}

\newcommand{\exlabel}[1]{\label{ex:#1}}
\newcommand{\exref}[1]{Example~\ref{ex:#1}}

\newcommand{\seclabel}[1]{\label{sec:#1}}

\newcommand{\longsecref}[1]{Section~\ref{sec:#1}}

\newcommand{\figlabel}[1]{\label{fig:#1}}
\newcommand{\longfigref}[1]{Figure~\ref{fig:#1}}

\newcommand{\secref}{\longsecref}
\newcommand{\figref}{\longfigref}

\numberwithin{equation}{section}

\definecolor{dkgreen}{rgb}{0,0.3,0}
\definecolor{gray}{rgb}{0.5,0.5,0.5}
\definecolor{mauve}{rgb}{0.58,0,0.82}
\definecolor{light-gray}{gray}{0.80}

\usepackage{listings}
\lstdefinelanguage{sketch}{
  morekeywords = {
       bool, harness, data, int, bit, bool, adt, float,new, return, assert , assume, let, case, switch, in, repeat\_case, generator, if, else, for },
  morecomment=[s]{/*}{*/},
}

\renewcommand{\scriptsize}{\fontsize{8.5}{9}\selectfont}
\makeatletter
\lst@AddToHook{TextStyle}{\let\lst@basicstyle\scriptsize\fontfamily{lmss}\selectfont}

\usepackage{tikz}
\usepackage{tikzscale}
\makeatother
\usetikzlibrary{shapes,arrows}
\tikzstyle{decision} = [diamond, draw, fill=blue!20, 
    text width=4.5em, text badly centered, node distance=3cm, inner sep=0pt]
\tikzstyle{block} = [rectangle, draw, fill=white!20, node distance = 6cm, text centered, rounded corners, minimum height=2em]
\tikzstyle{line} = [draw, -latex']
\tikzstyle{arrow}=[->, shorten >=1pt]
\tikzstyle{cloud} = [draw, ellipse,fill=red!20, node distance=5cm,
    minimum height=2em]
\tikzstyle{outer} = [rectangle, draw, node distance = 6cm, text centered, rounded corners, minimum height=2em]

\usepackage{url}

\usepackage{rotating}

\usepackage{enumitem}

\usetikzlibrary{positioning}
\usetikzlibrary{shapes,arrows}
\tikzstyle{decision} = [diamond, draw, fill=blue!20, 
    text width=4.5em, text badly centered, node distance=3cm, inner sep=0pt]
\tikzstyle{block} = [rectangle, draw, fill=white!20, node distance = 6cm, text centered, rounded corners, minimum height=2em]
\tikzstyle{line} = [draw, -latex']
\tikzstyle{cloud} = [draw, ellipse,fill=red!20, node distance=5cm,
    minimum height=2em]

%% file: commands.tex
\newcommand{\E}{E}
\newcommand{\B}{B}
\newcommand{\rhole}{x}
\newcommand{\bhole}{y}

\newcommand{\const}{c}

\newcommand{\sem}[2]{\llbracket #1 \rrbracket_{#2}}
\newcommand{\ctrls}{\sigma}

\newcommand{\LLN}{\alpha_N}
\newcommand{\LLB}{\alpha_B}

%% file: abstract.tex
In this paper, we present \tech{}, a technique that combines numerical optimization with SAT solving to synthesize unknowns in a program that involves discrete and floating-point computation. \tech{} makes the program end-to-end differentiable by \emph{smoothing} any Boolean expression that introduces discontinuity such as conditionals and \emph{relaxing} the Boolean unknowns so that numerical optimization can be performed. On top of this, \tech{} uses a SAT solver to help the numerical search overcome local solutions by incrementally fixing values to the Boolean expressions. We evaluated the approach on 5 case studies involving hybrid systems and show that \tech{} can synthesize programs that could not be solved by previous SMT approaches. 

%% file: intro.tex
\section{Introduction}
Gradient-based numerical techniques 
are becoming a popular mechanism for solving
program synthesis problems. Neural networks have shown that by doing automatic differentiation over complex computational structures (deep networks), they can solve many complex, real world problems~\cite{hinton2012deep, krizhevsky2012imagenet, silver2016mastering}.  
There is also a growing body of work on using neural networks to learn programs with discrete control structure from examples, such as neural Turing machines~\cite{graves2014neural} and others that follow similar ideas~\cite{reed2015neural, kurach2015neural, neelakantan2015neural, kaiser2015neural}. For many problems with discrete structure, however, recent work~\cite{gaunt2016terpret} has shown that neural networks are not as effective as state-of-the-art program synthesis tools like \Sk{}~\cite{solar2006combinatorial, sketchthesis} that is based on SAT and SMT solving.

This paper shows that a combination of SAT and gradient-based numerical optimization can be effective in solving synthesis problems that involve both discrete and floating-point computation.
The combination of numerical techniques and SAT is not itself new; SMT solvers such as Z3~\cite{z3},  dReal~\cite{gao2013dreal} and the more recent work on satisfiability modulo convex optimization (SMC) ~\cite{smc} have also explored problems involving the combination of discrete structure and continuous functions using the DPLL(T) framework~\cite{dpllt}. However, we show that the approach followed by SMT solvers, while effective for many applications, is sub-optimal for program synthesis problems. The key problem with prior approaches is the way in which they separate the discrete and the continuous parts of the problem, which loses high-level structure that could be exploited by gradient descent. A consequence of this loss of structure is that for some problems the SMT solver requires an exponential number of calls to the numerical solver; SMC mitigates this by focusing on a special class of problems called monotone SMC formulas but does not generalize to arbitrary problems. 

Our technique, called \tech{} (for Real Synthesis), exploits the full program structure by making the program end-to-end differentiable. It leverages automatic differentiation~\cite{autodiff} to perform numerical optimization over a smooth approximation of the \emph{full} program. At the same time, \tech{} uses a SAT solver to both deal with constraints on discrete variables and  
    constrain the search space for numerical optimization by fixing values of Boolean expressions in the program. This allows \tech{} to explore the numerical search space based on the structure of the program.


End-to-end differentiability is achieved by \emph{smoothing} the Boolean structure such as conditionals using a technique similar to~\cite{smooth} and \emph{relaxing} Boolean unknowns to reals in the range $[0,1]$ similar to mixed integer programming. 
The smoothing algorithm we use is a simplified version to what is used in~\cite{smooth}, replacing sharp transitions in conditionals with smooth transition functions such as \C{sigmoid}. Despite using a simpler smoothing approach, our technique works better for two reasons. First, the simpler smoothing approach allows us to use automatic/algorithmic differentiation, unlike~\cite{smooth} which had to rely on gradient free optimization techniques (Nelder-Mead) which are not as effective as the gradient-based methods. Most importantly, though, \sys{}'s use of a SAT solver
to fix the values of the Boolean expressions allows it to better tolerate the inevitable approximation error at branches, allowing us to get better results while using less precise (and more efficient) approximations compared to~\cite{smooth}. 
\noindent
\paragraph{The full system} We implemented the \tech{} technique  with the \Sk{} system as the front-end. Similar to \Sk{}, in \sys{} the programmer writes a high-level implementation  with unknowns. In our case, these unknowns can be either reals or Booleans. In addition, the programmer can also introduce assertions to 
specify the intended program behavior. The synthesis problem is to find values for these unknowns such that all the assertions in the program are satisfied. The front-end language of \Sk{} is very expressive which includes support for arrays, ADTs, heap-allocated structures, etc, and hence, all these features can be used to specify the problem in \sys{} as well.
\noindent
\paragraph{Results}
We use  \sys{} to solve some interesting synthesis problems that are more complex than anything that has been solved by prior work. In particular, we focus on synthesizing parametric controllers for hybrid systems. 
This is a good fit for \sys{}  because of the combination of continuous and discrete reasoning, and because the goal is to find satisfying assignments corresponding to the unknown parameters, as opposed to proving unsatisfiability, which our system is unable to do. 
We show that \sys{} can solve these problems in 3-11 minutes whereas previous SMT solvers such as dReal and Z3 cannot solve them. 
\noindent
\paragraph{Summary of Contributions}
\begin{itemize}
	\item We present \tech{}, a novel  technique to combine numerical optimization with SAT reasoning that allows us to do efficient reasoning on programs involving both discrete and continuous functions. 
	\item \tech{} achieves end-to-end differentiability in the presence of discrete structure using \emph{smoothing} and \emph{relaxing} techniques and it uses a SAT solver to make discrete decisions to overcome local solutions from numerical search. 
	\item We evaluated the system by synthesizing parametric controllers for several interesting hybrid system scenarios that previous systems cannot solve.
\end{itemize}

%% file: overview.tex
\section{Overview}\seclabel{overview}
In this section, we first present a stylized synthesis task to illustrate the kinds of problems \sys{} can handle and then use a synthetic example to illustrate how \sys{} works.

\subsection{Illustrative Example}

\begin{example}[Lane change controller]\exlabel{lanechange}
Consider the scenario shown in \figref{lanechange}. The task is to synthesize a controller program that can move the car 1 from lane 1 to lane 2 within $T$ time-steps  without colliding with any of the other cars. 
\end{example}
\noindent
The program with \emph{unknowns} for the  controller is shown in \figref{lanechange_template}. The controller is composed of 5 modes. Each mode has rules for setting the two controls of the car--the acceleration and the steering angle. The synthesizer must discover a sequence of steps to perform the lane change; first accelerating until it is in the correct position to do the lane change, then starting the lane change by turning the wheels to the left, then turning right once it is in the new lane, then adjusting its velocity to match the other cars, and finally maintaining a stable velocity.
The unknowns in the program are the switching conditions for changing from one mode to another, and the exact values for the acceleration and the steering angle in each mode. 

The space of switching conditions is described by the function \C{genSwitchExp()} which encodes a discrete choice between four different inequalities on the relevant state variables. Note that the \C{genSwitchExp} function is marked as a \emph{generator}. A generator in \tech{} is treated as a macro that gets inlined at every call, and the solver can choose different values for the unknowns for each different instantiation. 

 \figref{lanechange_spec} shows the specification for this synthesis task.  The specification simulates $T = 50$ time steps where at each time step, it calls the controller code that sets the control values based on the current state of the cars and then, moves the car and the world by one time step ($dt = 0.1s$). In this problem, we assume that the other vehicles around Car 1 in the world have a constant velocity. Finally, the specification asserts that there is no collision at any time step and the goal is reached at the end of the simulation. 

\begin{figure}
	\includegraphics[width=0.40\textwidth]{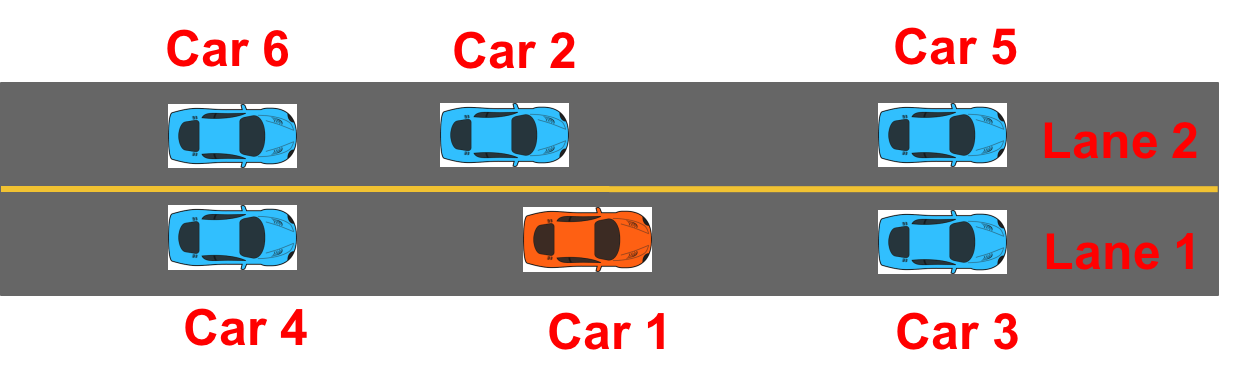}
	\caption{Lane change initial scenario}
	\figlabel{lanechange}
\end{figure}

\begin{figure}
	\begin{lstlisting}
	void LaneChangeController (Car car1) {
		if (genSwitchExp(car1)) { /* go straight */
			car1.a = ??%*$_r$*);
			car1.%*$\theta$*) = 0; 
		} else if (genSwitchExp(car1)) { /* turn left */
			car1.a = ??%*$_r$*);
			car1.%*$\theta$*) = ??%*$_r$*);
		} else if (genSwitchExp(car1)) { /* turn right */
			car1.a = ??%*$_r$*);
			car1.%*$\theta$*) = ??%*$_r$*);
		} else if (genSwitchExp(car1)) { /* go straight */
			car1.a = ??%*$_r$*);
			car1.%*$\theta$*) = 0;
		} else { /* rest state */
			car1.a = 0;
			car1.%*$\theta$*) = 0;	}}
	generator bool genSwitchExp(Car car1) {
		if (??%*$_b$*)) return car1.x %*$\leq$ *) ??%*$_r$*);
		else if (??%*$_b$*)) return car1.x %*$\geq$ *) ??%*$_r$*);
		else if (??%*$_b$*)) return car1.y %*$\leq$*)  ??%*$_r$*);
		else return car1.y %*$\geq$*) ??%*$_r$*);	}
	\end{lstlisting}	
	\caption{Template for the lane change controller.}
	\figlabel{lanechange_template}
\end{figure}

\begin{figure}
	\begin{lstlisting}
	LaneChangeSpec (Car car1, World w) {
		for (int i = 0; i < 50; i++) {
			LaneChangeController(car1);
			MoveCar(car1);
			MoveWorld(w);
			assert (!DetectCollision(car1, w));
		}
		assert (ReachedGoal(car1, w));	}
	\end{lstlisting}
	\caption{Specification for the lane change synthesis task.}
	\figlabel{lanechange_spec}
\end{figure}

Solving this problem is  a significant challenge for three reasons. First, the five-mode controller program is simulated 50 times in the specification, so there are $8\times10^{34}$  paths in the synthesized program. Moreover 
at each timestep, \C{DetectCollision} has to check for collisions against each of 5 other cars, and each of these checks has to consider 8 separate conditions that can indicate a collision between two cars, so there are a total of 2000 checks. Second, this sketch has 12 Boolean unknowns and 22 real unknowns, so there is a very large space to search. 
Finally, we use a bicycle-model for the dynamics of the car. Even though this model is simpler than dynamics of a real car, it is still fairly complex and involves non-linear functions such as sine, cosine, and square root.

Because of the these challenges, existing approaches do not perform well on this synthesis task. Many SMT solvers such as Z3 do not provide full support for non-linear real arithmetic and hence, cannot synthesize this program. We also found that dReal, an SMT solver for reals, is unable to solve this problem (see \secref{eval}). 
On the other hand, smoothing approaches also fail because of the amount of boolean structure in the problem. For comparison, the most complex problem that was solved by the system in~\cite{smooth} was similar to this one, but involved only 1 obstacle (not 5), with 2 collision conditions (instead of 8), and only scaled to 35 time steps. The use of automatic differentiation helps, but in \secref{eval}, we show that for this benchmark, a smoothing approach with automatic differentiation cannot find a solution even with 300 trials from random initial points (which took about 40 minutes). However, despite all these difficulties, \tech{} can synthesize a correct program in  11 minutes.

\subsection{The \tech{} Approach}
We, now, describe the key ideas of the algorithm in the context of the example below. The example looks contrived because it was engineered to highlight all the key features of the algorithm.
\begin{example}  \exlabel{simpleex}
~
\begin{lstlisting}
	float x%*$_1$*) = ??%*$_r$*);
	assert(-20 %*$\le$*) x%*$_1$*) %*$\le$*) 6);
	float a = x%*$_1$*)-5;
	if (x%*$_1$*) %*$\le$*) 4)  a = 6- x%*$_1$*);
	if (x%*$_1$*) %*$\le$*) 2)  a = 8- x%*$_1$*);
	if (x%*$_1$*) %*$\le$*) 0)  a = 21+ x%*$_1$*);
	assert(a %*$\le$*) 0 || a  > 25);
\end{lstlisting}

\end{example}

\paragraph{SMT Solving Background. }

To understand why an SMT solver would do a suboptimal job solving for a value of $x_1$ in the program above, it is important to understand how an SMT solver works. As a first step, the program above would be converted into a logical formula that would then be separated into a boolean skeleton and a conjunction of constraints in a theory. 

In the example above, the solver would generate boolean variables corresponding to each of the constraints in the theory. 
\[
\begin{array}{lll}
y_1 = x_1\leq 0  & y_2 = x_1\leq 2 & y_3 =  x_1\leq 4\\
y_4 = a \leq 0 &  y_5 = a > 25 & y_6 = -20 \leq x_1 \leq 6\\
t_1 = (a_0 = x_1-5) & t_2 = (a_1 = 6-x_1) & t_3 = (a_2 = 8-x_1)  \\
t_4 = (a_3=21+x_1) & t_5 = (a_4=a_1) & t_6 = (a_4 =a_0)  \\
 t_7 = (a_5=a_2) & t_8 = (a_5=a_4) &  t_9 = (a=a_3)  \\
t_{10} = (a=a_5) & & \\
\end{array}
\]
The names $a_0$ to $a_5$ correspond to the temporary values of $a$ at each step of the computation. 
The boolean constraints include constraints corresponding to the 
initial assertions in the program (just $y_6 \wedge (y_4 \vee y_5)$), 
as well as constraints that describe the control flow, shown below:
\[
\begin{array}{l}
t_1 \wedge t_2 \wedge t_3 \wedge t_4 \wedge 
(y_3 \Rightarrow t_5) \wedge (\overline{y_3} \Rightarrow t_6) \\
(y_2 \Rightarrow t_7) \wedge (\overline{y_2} \Rightarrow t_8) \wedge
(y_1 \Rightarrow t_9) \wedge (\overline{y_1} \Rightarrow t_{10}) 
\end{array}
\]
Breaking the problem in this manner allows for a clean separation between boolean reasoning which is the responsibility of a SAT solver and theory reasoning, but it deprives the theory solver for crucial information about the control flow structure of the program. In the worst case, the SMT solver has to invoke the theory solver once for every path in the program. Lazy SMT solvers deploy many strategies to avoid the exponential number of calls. In the example above, a good solver solver would be able to generate theory propagation lemmas that show, for example, that $y_1$ implies $y_2$,
which would prevent it from considering infeasible paths, 
but even then, in the worst case the solver would still have to invoke the theory solver for each of the cases $x_1\in [-20, 0], (0, 2], (2, 4], (4, 6]$. When you have complex non-linear arithmetic, generating those theory propagation lemmas is much more challenging. For example, if condition $y_1$ were instead $x_1^3+8x_1^2-28x_1 < 80$, the program would be semantically equivalent to the one above, but it would be much harder for an SMT solver to avoid having to perform exponentially many calls to the theory solver.

The problem is even worse for a program like the one in \exref{lanechange}, with its $10^{34}$ possible paths and with complex non-linear relationships between the branches in different iterations.

\paragraph{Numerical optimization with Smoothing. }

The first important feature of \tech{} is the ability to use numerical optimization by performing automatic differentiation on a smooth approximation of the program. \figref{exgraph}(a) shows the graph for $a$ as function of $x_1$ and \figref{exgraph}(b) shows its smooth approximation. What jumps out from the graph is that while the branches do introduce discontinuities, the function is still very amenable to numerical optimization. However, numerical optimization can introduce its own problems. For example, even with the ability to smooth away discontinuities and automatically compute derivatives for the whole program, it is clear from the figure that gradient-based optimization will only succeed if it starts at $x_1 > 0$. Otherwise, the algorithm will be stuck on a local minima. In this example, initializing numerical search uniformly at random on the allowed range of $x_1$ would give a $77\%$ probability of failing with a local minima, which is already bad, but it is not hard to see how a small change to the program could make this probability arbitrarily close to 100\%. Smoothing can ameliorate the problems inherent in numerical search, but cannot eliminate them altogether. 

\paragraph{Using SAT solver to eliminate local solutions.}

The \tech{} approach is to turn the SMT paradigm on its head. In SMT, the SAT solver always has an abstraction of the complete problem. The theory solver helps refine this abstraction, and checks candidate assignments for consistency with the theory, but the SAT solver is the one driving the process. In \tech{}, the numerical solver is the one that drives the process. 
In the beginning, the numerical solver has a smooth approximation of the entire program and uses automatic differentiation and numerical optimization to find a local optima. In the case of the example, that first iteration of gradient descent is likely to converge to the local optima at $x=-20$ which fails to satisfy the constraint. 
When this happens, \tech{} asks the SAT solver for a boolean assignment that is used to guide the search; for example, 
the SAT solver may suggest setting $y_1=x_1\leq 0$ to $false$. At this point, the numerical solver performs a new round of numerical optimization, but now under the assumption that $x_1 > 0$, and therefore with one fewer branch compared to the previous case. 
In this case, setting this boolean condition is sufficient to steer the numerical optimization to a region where it can converge to a value that satisfies the constraint.

Once the numerical solver finds a solution that satisfies the constraints, say $x=4.01$, it still needs to check that it really satisfies the boolean constraints and that it is a true solution and not an artifact of the smoothing transformation. It does this by suggesting assignments to the SAT solver corresponding to that solution; in this case $\overline{y_1}, \overline{y_2}, \overline{y_3}, y_4, \overline{y_5}$. As the SAT solver sets these variables and checks them against the boolean constraints, the numerical solver checks that the current solution is still valid when the respective branches have been fixed in the program. As a result, the SAT solver helps refine the solution provided by the numerical solver, until a precise solution to the overall problem has been produced.

\begin{figure}
\begin{tabular}{cc}
\includegraphics[width=0.24\textwidth]{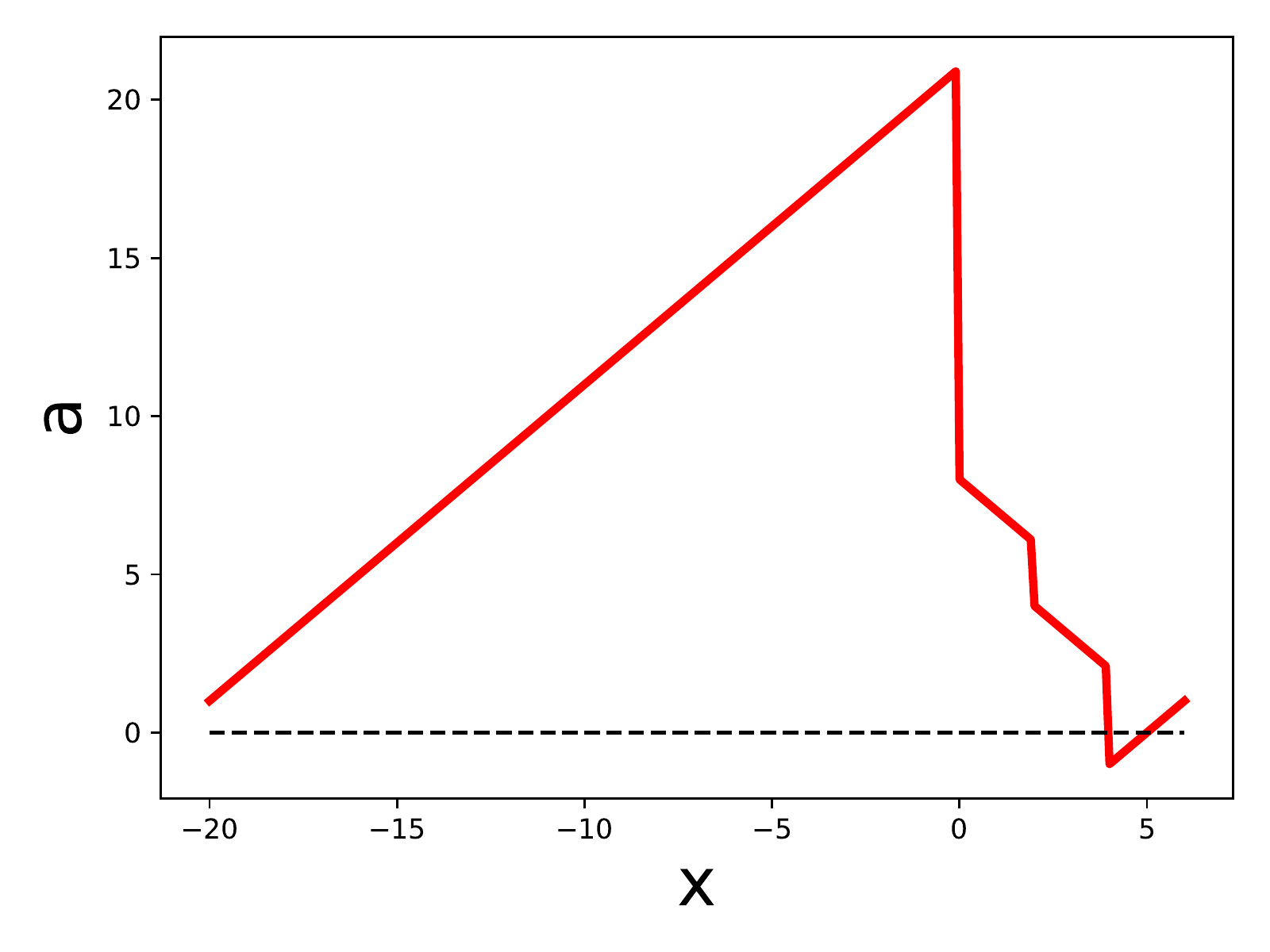} &
\includegraphics[width=0.24\textwidth]{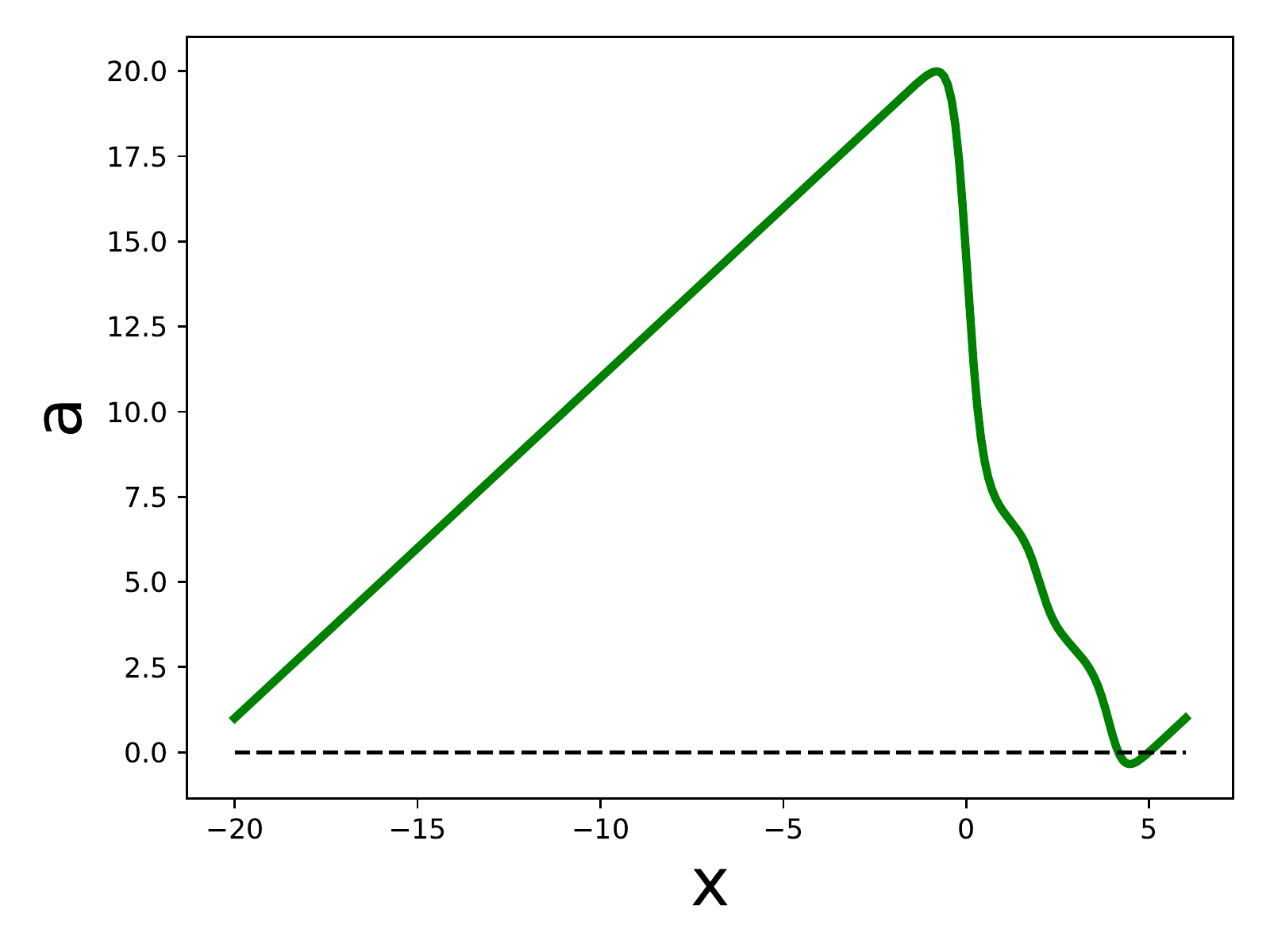}\\
(a) & (b)
\end{tabular}
\caption{(a) Graph showing $a$ vs $x_1$ in \exref{simpleex}. (b) Its smooth approximation.}
\figlabel{exgraph}
\end{figure}

%% file: synthesis.tex
\section{The \tech{} Intermediate Representation}
The core language (L) of \sys{} is shown below. The language consists of real-valued expressions $E$, Boolean expressions $B$ and Boolean unknowns $H$. The real-valued expressions can either be a real unknown $\rhole$, a constant $\const$, a real-valued operation \C{op} such as addition, multiplication, sine, cosine, etc, or a if-then-else expression \C{ite}. The Boolean expressions can either be a comparison  $E \ge 0$,  a conjunction,  or a negation. There are two kinds of \C{ite} expressions in the language--one where the conditional is a Boolean expression and the other where the conditional is a Boolean unknown because they are treated differently for the numerical problem. The language  does not allow Boolean unknowns to appear anywhere other than as a conditional. However, this does not decrease the expressive power of the language since those cases can be reduced to the core language easily. The imperative programs shown in \secref{overview} can be converted into this core language using straight-forward transformation passes and loop unrolling.  
\begin{eqnarray*}
	E & := & \rhole \; \vert \; \const \; \vert \; \t{op}(\{E_i\}_i) \; \vert \; \t{ite}(B, E_1, E_2) \; \vert \; \t{ite}(H, E_1, E_2)\\
	B & := &   E \ge 0 \; \vert \; B_1 \wedge B_2 \; \vert \; \neg B \\
	H & := & \bhole
\end{eqnarray*}


The semantics of some features of the language is shown below. It is defined in terms of a mapping $\ctrls$ from the unknowns to actual values. 
\begin{eqnarray*}
	\sem{\rhole}{\ctrls} & := & \ctrls[\rhole]	\\
	\sem{\bhole}{\ctrls} & := & \ctrls[\bhole]  \hspace{10mm}// ~\{0,1\} \\
	\sem{\E_1 + E_2}{\ctrls} & := & \sem{E_1}{\ctrls} + \sem{E_2}{\ctrls}\\
	\sem{\t{ite}(B, E_1, E_2)}{\ctrls} & := & \sem{B}{\ctrls} . \sem{E_1}{\ctrls} + \sem{\neg B}{\ctrls} .  \sem{E_2}{\ctrls}\\
	\sem{E \ge 0}{\ctrls} & := & \sem{E}{\ctrls} \ge 0 
\end{eqnarray*}

\subsection{Synthesis Problem}
The synthesis problem is given a program $P = \t{assert}(B_1, \cdots, B_k)$, find $\ctrls$ such that $\forall k. \sem{B_k}{\ctrls}  = 1$. For now on, we will use $\psi(\ctrls)$  to represent this predicate.  So, the synthesis problem is to solve the formula: $\exists \ctrls.~\psi(\sigma)$.

\sys{}  divides the synthesis problem into a SAT part ($\psi_B$) and a numerical part ($\psi_N$).  We, first, describe the abstraction process for producing $\psi_B$ and $\psi_N$ and \secref{solver} describes the algorithm to solve $\psi$ by repetitively solving $\psi_B$ and $\psi_N$.

\subsection{Boolean Abstraction ($\alpha_B$)}

\newcommand{\Babs}{\widetilde{B}}
\newcommand{\PBabs}{\widetilde{P_B}}
\newcommand{\match}{\textsc{MatchNUpdate}}
The Boolean abstraction is obtained similar to the process used  in SMT solvers. Concretely, the language for Boolean constraints ($L_B$) is:
\begin{eqnarray*} 
	\PBabs & := & \t{assert}(\Babs_1, \cdots, \Babs_k) \\
	\Babs & := & \bhole \; \vert \; \Babs_1 \wedge \Babs_2 \; \vert \; \neg \Babs
\end{eqnarray*}
 The language is almost same as the $B$ expressions in the original language. The only exception is that the term $E \ge 0$ is now replaced by  Boolean unknowns. 

The Boolean abstraction, $\LLB$, is a function from $L$ to $L_B$ and is defined as below: 
\begin{eqnarray*}
	\LLB(\bhole) & = & \bhole \\
	\LLB(E \ge 0) & = & \t{Create new } \bhole \\
	\LLB(B_1 \wedge B_2) & = & \LLB(B_1) \wedge \LLB(B_2)\\
	\LLB(\neg B) & = & \neg \LLB(B)
\end{eqnarray*}

\subsection{Numerical Abstraction ($\alpha_N$)}
\newcommand{\Eabs}{\widetilde{E}}
\newcommand{\PEabs}{\widetilde{P_E}}
\newcommand{\I}{I}
\newcommand{\Iname}{interface mapping}
We now describe the abstraction process to produce $\psi_N$ from $\psi$. The abstraction is defined with respect to a function $\I$, called \emph{\Iname{}}, that maps every Boolean expression in $\psi$ to one of $\{0, 1, \bot\}$. $\I(B_i) = v_i \neq \bot$ means that the Boolean expression $B_i$ should have the value $v_i$. If $v_i = \bot$, then the value is not yet set. 
The interface mapping specializes a synthesis problem defined by $\psi$ to $$\exists \ctrls.~ \psi[\{B_k/v_k\}_k](\ctrls) \wedge_i \sem{B_i[\{B_k/v_k\}_{k\neq i}]}{\ctrls} = v_i$$
where the spec $\psi$ is first simplified by substituting the expressions $B_k$ with the values $v_k=\I(B_k)$ when $v_k\neq \bot$, and additional constraints are added to ensure that $\ctrls$ satisfies the assignments in $\I$. 

The goal of the numerical abstraction ($\alpha_N$) is to produce a \emph{smooth}   approximation $\psi_N$ of the specialized problem above that can be fed to an off-the-shelf numerical solver to find an assignment to $\sigma$ if one exists. Note that unlike the SMC approach,  \tech{}  does not require $\psi_N$ to be composed of linear/convex functions.  However, we want $\psi_N$ to be smooth and continuous because numerical algorithms  perform poorly in the presence of discontinuities. The main source of discontinuity arises when the Boolean expression in an \C{ite} or a conjunction is not yet set by $\I$. \tech{} eliminates these discontinuities by performing a program transformation that replaces the sharp transitions with smooth transition functions such as \C{sigmoid} as described in the next subsection.

\newcommand{\p}{p}
\newcommand{\e}{e}
\newcommand{\enew}{\widetilde{e}}
\newcommand{\be}{b}
\newcommand{\fs}{\mathcal{F}_s}

\subsubsection{Abstraction rules}
The language for numerical constraints ($L_N$) in \sys{} is shown below. At the top level, we have a conjunction of numerical inequalities $(\Eabs_i \ge 0)$ where $\Eabs$ is again similar to $E$ in the original language minus the \C{ite} expression. 
\begin{eqnarray*}
	\PEabs & := &  \Eabs_1 \ge 0 \wedge \cdots \wedge \Eabs_n \ge 0\\
	\Eabs & := & \rhole  \; \vert \; \const \; \vert \; \t{op}(\{\Eabs_i\}_i) 
\end{eqnarray*}
Note that there are numerical algorithms (such as sequential quadratic programming) that take in a conjunction of  inequalities and perform constrained optimization on them directly.  Even for numerical algorithms that can only perform unconstrained optimization (such as plain gradient descent), it is  easy to transform a conjunction of inequalities into a smooth objective function.

Given a program $P = \t{assert}(B_1, \cdots, B_k)$ in $L$, the goal of the abstraction is to produce a program $\PEabs$ in $L_N$. In order to do that, we define a transformation rule $\e \xrightarrow{\I} (\enew, \p)$ where $\e$ is an expression in $L$ (either an $E$ or a $B$ expression), $\enew$ is the corresponding expression in $L_N$, and  $\p$ is the conjunction of numerical constraints obtained so far. This formulation allows us to collect numerical constraints from intermediate expressions if necessary. The transformation is defined in terms of two parameters: $(\beta, \epsilon)$. $\beta$ is the smoothing parameter that controls how smooth the approximation should be. Higher values for $\beta$ mean less smoothing. $\epsilon$, a small positive constant, is the precision of numerical calculation that arises due to im-precise floating point computation. Because of this, the expression $\Eabs \ge 0$ in $L_N$ actually means $\Eabs \ge -\epsilon$.

The rules for performing the abstraction for expressions in the language are shown in \figref{abs_algorithm}. We first focus on the  rules for expressions other than \C{ite} and conjunction expressions.  For an $E$ expression, the abstraction produces an $\Eabs$ that smoothly approximates the original expression. For simple cases such as a real unknown and constant, the abstraction just returns the same expression as shown in the \C{RHOLE} and \C{CONST} rules. In this case, there are no intermediate constraints and hence, $\p$ is just true (\C{T}). For an \C{op} expression, the \C{OP} rule recursively smooths the expressions in its arguments and then creates a new \C{op} expression with these smoothed replacements and concatenates the intermediate constraints obtained from abstracting the arguments. \sys{} assumes that \C{op} is itself a smooth and continuous operation. For operations like division or $sqrt$ that are not defined for all inputs, we replace them with continuous approximations, and rely on the frontend to introduce assertions that ensure their inputs stay away from the regions where the approximations differ significantly from the true operations.

\newcommand{\dist}{P-distance}
\newcommand{\distf}{{\mathcal{P}_d}}

In order to understand the abstraction rules for the $B$ expressions, we first define a function called \dist{}.
\begin{definition}[\dist{}]
	The \dist{} (short for positive-distance) for a Boolean expression $\be$ in $L$ is a function $\distf$ that takes in an assignment to the unknowns $\ctrls$ and produces a real value such that $\forall \ctrls.~(\distf(\ctrls) \ge \epsilon \implies \sem{\be}{\ctrls} = 1) \wedge (\distf(\ctrls) \le -\epsilon \implies \sem{\be}{\ctrls} = 0)$.  Thus, for numerical purposes, we can assume $b \approx \distf \ge 0$. 
    There can be more than one \dist{} function for the same Boolean expression. 
\end{definition}
The abstraction rule for a $\B$ expression in $L$ produces an $\Eabs$ expression that is a smooth approximation to a \dist{} function for $\B$. For example, for  $\e \ge 0$, the natural choice for its \dist{} function is $\e$ itself. So, the rule \C{GE} first computes the smooth approximation of $\e$. For Boolean expressions, the rules also need to take into account whether there is a value assigned to them in the $\I$ mapping. This is done using the \match{} function which takes in the smoothed expression $\enew$ and the value $v$ from $\I$. Then, it checks for one of the three cases: 1. if $v = \bot$, it means that the value is not yet set and hence, there is no update and no new constraints, 2. if $v = 1$, then the expression $\enew$ is replaced with a large positive constant ($K = 100$) and a new constraint $\enew \ge 0$ is added, and 3. if $v = 0$, then similar to case 2, the expression $\enew$ is replaced with a large negative constant and the condition $-\enew \ge 0$ is added~\footnote{Note that the actual condition in this case should be $-\enew > 0$, but because of the floating-point precision issue, we write it as $-\enew \ge0$}. The reason for replacing $\enew$ with constants in cases 2 and 3 is so that other expressions that depend on $\enew$ can infer the Boolean value of the $\B$ expression it represents without any ambiguity (because of the large magnitude of these constants).
The \C{NOT} rule for the $\neg \be$ expression, similarly, first computes the abstraction for $\be$ and negates the expression obtained to get a smooth approximation to a \dist{} for $\neg \be$. 

Finally, the abstraction for the assert expression iteratively smooths each of its $\be$ arguments and  creates the final set of numerical constraints $\PEabs$ to form the numerical problem $\psi_N$. This final set of constraints is the conjunction of the constraints obtained after transforming each Boolean argument ($\p_i$) and as well as constraints to ensure that the abstraction of each Boolean argument ($\enew_i$) is greater than $0$.

\begin{example}
	Consider the program $P = \t{assert}(\t{ite}(x_1 \ge 0, x_2, x_3) \ge 0)$ where $x_1, x_2, x_3$ are real unknowns and let $\I(x_1 \ge 0) = 0$. The abstraction for $x_1 \ge 0$ results in $(-100, -x_1 \ge 0)$. So, the \C{ite} expression can be thought  as $\t{ite}(-100 \ge 0, x_2, x_3)$. Even though we have not yet discussed the rule for abstracting  \C{ite} expressions, in this case, it is clear that the abstraction should just produce $x_3$. Overall, abstracting $P$, in this example, will result in $\PEabs = x_3 \ge 0 \wedge -x_1 \ge 0$. 
\end{example}

\begin{figure}[t!]
	\small
	\begin{eqnarray*}
	&	\t{ASSERT} ~~~ \frac{\begin{array}{c} 
				\be_1 ~~\xrightarrow{\I}~~(\enew_1, \p_1) ~\cdots ~ \be_k ~~ \xrightarrow{\I}~~(\enew_k, \p_k)\\
				\p = \p_1 \wedge \cdots \wedge \p_k \wedge (\enew_1 \ge 0) \cdots \wedge (\enew_k \ge 0)
			\end{array}} 
			{\begin{array}{c}\t{assert}(\be_1, \cdots, \be_k)~~ \xrightarrow{\I}~~ (0, \p) \end{array}}\\
	& \t{GE}~~~\frac{\begin{array}{c}
	 			\e ~~\xrightarrow{\I}~~(\enew, \p)\\
	 			 (\enew_\I, \p_\I) = \match(\enew, \I(\e \ge 0)) \\
	 		\end{array}}
 			{\begin{array}{c}\e \ge 0~~\xrightarrow{\I}~~(\enew_\I, \p \wedge \p_\I) \end{array}}\\
 	& \t{NOT}~~~\frac{\begin{array}{c}
 			\be ~~ \xrightarrow{\I}~~(\enew, \p)\\
 			(\enew_\I, \p_\I) = \match(-\enew, \I(\neg \be))
 			\end{array}}
 			{\begin{array}{c} \neg \be ~~ \xrightarrow{\I}~~(\enew_\I, \p \wedge \p_\I)  \end{array}}\\
 	& \t{AND}~~~\frac{\begin{array}{c}
 			\be_1 ~~\xrightarrow{\I}~~(\enew_1, \p_1) \qquad \be_2 ~~\xrightarrow{\I}~~(\enew_2, \p_2) \\
 			\enew = \enew_1*t + \enew_2*(1 - t),~~ t = \fs(\enew_2 - \enew_1)\\
 			(\enew_\I, \p_\I) = \match(\enew, \I(\be_1 \wedge \be_2))
 	\end{array}}
 	{\begin{array}{c} \be_1 \wedge \be_2 ~~\xrightarrow{\I}~~(\enew_\I, \p_1 \wedge \p_2 \wedge \p_\I) \end{array}}\\
 	& \t{RHOLE}~~~\frac{\begin{array}{c}
 			~
 			\end{array}}
 		{\begin{array}{c} x  ~~ \xrightarrow{\I}~~(x, \C{T}) \end{array}}\\
 	& \t{CONST}~~~\frac{\begin{array}{c}
 			~
 	\end{array}}
 	{\begin{array}{c} c  ~~ \xrightarrow{\I}~~(c, \C{T}) \end{array}}\\
 	& \t{OP}~~~\frac{\begin{array}{c}
 			\e_i ~~\xrightarrow{\I}~~(\enew_i, \p_i)
 	\end{array}}
 	{\begin{array}{c} \t{op}(\{\e_i\}_i)  ~~ \xrightarrow{\I}~~(\t{op}(\{\enew_i\}_i), \underset{i}{\wedge}\p_i) \end{array}}\\
 	& \t{ITE1} ~~~ \frac{\begin{array}{c}
 			\e_1 ~~\xrightarrow{\I}~~(\enew_1, \p_1) \qquad \e_2 ~~ \xrightarrow{\I} ~~ (\enew_2, \p_2)\\
 			\be ~~ \xrightarrow{I} ~~(\enew_c, \p_c)\\
 			\enew = \enew_1*\fs(\enew_c) + \enew_2*(1- \fs(\enew_c))
 			\end{array}}
 		{\begin{array}{c} \t{ite}(\be, \e_1, \e_2) ~~ \xrightarrow{\I} ~~ (\enew, \p_1 \wedge \p_2 \wedge \p_c)	\end{array}}\\
 	& \t{BHOLE1} ~~~ \frac{\begin{array}{c}
 			v = \I(\bhole)  \neq \bot
 			\end{array}}
 		{\begin{array}{c} \bhole ~~ \xrightarrow{\I} ~~ (v, \C{T}) \end{array}}\\
 	& \t{BHOLE2} ~~~ \frac{\begin{array}{c}
 			\I(\bhole)  = \bot\\
 			x = \t{Create new real unknown for } \bhole\\
 			\p = (x \ge 0) \wedge (1 - x \ge 0)\\
 			\p = \p \wedge (\delta - x(1-x) \ge 0)
 	\end{array}}
 	{\begin{array}{c} \bhole ~~ \xrightarrow{\I} ~~ (x, \p) \end{array}}\\
 	& \t{ITE2} ~~~ \frac{\begin{array}{c}
 			\e_1 ~~\xrightarrow{I}~~(\enew_1, \p_1) \qquad \e_2 ~~ \xrightarrow{\I} ~~ (\enew_2, \p_2)\\
 			\bhole ~~\xrightarrow{\I} ~~ (\enew_c, \p_c)\\ 
 			\enew = \enew_1*\enew_c + \enew_2*(1 - \enew_c)\\
 			\end{array}}
 		{\begin{array}{c} \t{ite}(\bhole, \e_1, \e_2) ~~ \xrightarrow{\I} ~~ (\enew, \p_1 \wedge \p_2 \wedge \p_c) \end{array}}\\
 	& \match(\enew, v) = 
 		\begin{cases}
 			(\enew, \C{T}), & \t{if}\ v = \bot \\
 			(K, (\enew \ge 0)), & \t{if}\ v = 1 \\
 			(-K, (-\enew \ge 0)), & \t{if}\ v = 0
 		\end{cases}\\
 	\end{eqnarray*}
	\caption{Numerical abstraction rules. \C{T} stands for true, $K$ is a large positive constant (100) and $\delta$ is a small positive constant. Note that expressions such as $\e_1 * \e_2$ are  symbolic expressions.  $\fs$ is a smooth transition function.}
	\figlabel{abs_algorithm}
\end{figure}

Now, we can look at the rules for expressions that introduce discontinuities i.e. if-then-else and conjunctions.

\paragraph{If-then-else}
Let us first consider the \C{ite} expression of the form $\t{ite}(\be, \e_1, \e_2)$ that has a Boolean expression as the condition.  The rule \C{ITE1} describes the abstraction for these expressions. The rule first recursively smooths the expressions $\e_1$ and  $\e_2$ resulting in expressions $\enew_1$ and $\enew_2$. Then, the condition $\be$ is also transformed to produce the approximation for its \dist{} function $\enew_c$. If we were to use the actual semantics of the \C{ite} expression, we would get $\enew_1 * (\enew_c \ge 0) + \enew_2 * (1 - (\enew_c \ge 0))$. Note that the operations \C{+}, \C{*} in the above expression are actually symbolic operations. Clearly, the function $\enew_c \ge 0$ has a discontinuity at $\enew_c = 0$. To overcome this discontinuity, the \C{ITE1} rule replaces $\enew_c \ge 0$ with $\fs(\enew)$ where $\fs$ is a smooth transition function:
$$\fs(x) = \t{sigmoid}_{\beta}(x) = \frac{1}{1 + e^{-\beta x} }$$

The rule \C{ITE2} describes the abstraction for expressions of the form $\t{ite}(\bhole, \e_1, \e_2)$. In this case, $\bhole$ is abstracted using the \C{BHOLE1} or the \C{BHOLE2} rule. In the case where $\I$ already fixes the value of $\bhole$ to $1$  or $0$, the ite expression will be simplified to just $\enew_1$ or $\enew_2$ respectively. If the value of $\bhole$ is not yet set, then the rule \C{BHOLE2} creates a new real unknown $x$ corresponding to $\bhole$. This new variable is constrained to be in the interval $[0, 1]$. In addition, the constraint $x(1-x) \le \delta$ where $\delta = 0.1/\beta$ is  added to enforce that $x$ is either close to $0$ or $1$.  The \C{ite} expression is, then, abstracted by a linear combination of $\enew_1$ and $\enew_2$ such that when $x = 1$, the abstraction will result in $\enew_1$ and when $x = 0$, the result will be $\enew_2$. 

\paragraph{Conjunctions}
The rule for abstracting a conjunction of two Boolean expressions  is based on the following lemma:
\begin{lemma}
Let $\distf_1$ and $\distf_2$ be the \dist{} functions for  Boolean expressions $\be_1$ and $\be_2$. Then, $\distf = \t{min}(\distf_1, \distf_2)$ is a \dist{} function for $\be_1 \wedge \be_2$. 
\end{lemma}
Based on the lemma, the \C{AND} rule first gets the abstractions of $\be_1$ and $\be_2$ and then smooths the $\t{min}$ of the results. The smoothing is done by rewriting $\t{min}(\enew_1, \enew_2)$ as $\t{ite}(\enew_2 - \enew_1 \ge 0, \enew_1, \enew_2)$ and applying the \C{ITE1} rule. 

\subsubsection{Computing gradients}
One of the advantages of our numerical abstraction algorithm is that once the smoothed program $\PEabs$ is produced, we can use automatic differentiation~\cite{autodiff} to symbolically compute the gradients necessary to perform  gradient-based numerical search.  For example, consider the expression $\e = \e_1 * \e_2 + \e_3$ with two unknowns $\rhole = [\rhole_1, \rhole_2]$ and let $\ctrls$ be an assignment to the unknowns. Since there are two unknowns, each sub-expression will have two gradients, i.e. $\Gamma(e, \ctrls) = \left [\left ( \frac{\partial \e}{\partial \rhole_1} \right )_\ctrls, \left ( \frac{\partial \e}{\partial \rhole_2}\right )_\ctrls \right ]$ where $\Gamma$ is the notation used to get the gradients for any sub-expression $e$ at the assignment $\ctrls$. 
Automatic differentiation applies the chain rule repeatedly to each elementary expression i.e. in the above example  $$\Gamma(\e, \ctrls) = \sem{\e_2}{\ctrls} * \Gamma(\e_1, \ctrls) + \sem{\e_1}{\ctrls}* \Gamma(\e_2, \ctrls) + \Gamma(\e_3, \ctrls) $$
This process allows us to calculate the gradients accurately and  in time that is proportional to the time it takes to evaluate an expression.

\subsection{Properties of Numerical Abstraction}
Let $\e$ be the original expression, $\enew$  be the result of the abstraction and let $x$ be the set of unknown reals in $\enew$, then the following theorems hold for any $I$.

\begin{theorem}[Continuity]
	$\enew$  is continuous with respect to  $x$ under the assumption that all \C{op} operations in $\e$ are continuous with respect to their operands. 
\end{theorem}

\begin{theorem}[Differentiability]
	$\frac{\partial \enew}{\partial x}$ is continuous with respect to $x$ under the assumption that  all \C{op} operations in $\e$ are differentiable with respect to their operands. 
\end{theorem}

\begin{theorem}[Closeness]
$\underset{\beta \rightarrow \infty}{lim} \e \approx \enew$ 
\end{theorem}

This theorem states that when $\beta \rightarrow \infty$, both $\e$ and its abstraction $\enew$ agree on almost all  assignments to the unknowns. We say almost because $\e$ and $\enew$ may not agree at the branching points. For example, consider $\t{ite}(x \ge 0, 1, 0)$. In this case when $x = 0$, the abstraction will have undefined behavior.

%% file: solver.tex
\section{Numerical Solver + SAT Solver}\seclabel{solver}
Given a synthesis problem $\exists \ctrls.~\psi(\ctrls)$, the last section described the algorithms to produce the Boolean abstraction $\psi_B$  and the numerical abstraction $\psi_N$  when given an \Iname{} $\I$. In this section, we will describe the algorithm to combine the abstractions so that the original problem can be solved. 


\subsection{Architecture}
\begin{figure}
	\input{interface}
\end{figure}

The high-level architecture showing the interaction between the different components of the system is shown in \figref{overview}.
First, there is a \tech{} core that acts as an interface between the numerical solver and the SAT solver. The core is responsible for creating the appropriate abstractions for the numerical solver and the SAT solver and invoking them when necessary to solve $\psi$. The core also maintains the \Iname{} ($\I$). The core will be discussed in detail in \secref{core}

The numerical solver component can be any black box gradient-based algorithm that can take in a set of constraints $\psi_N^I$ and perform numerical search to produce SAT/UNSAT along with the satisfying assignment to the real unknowns. There can be two issues with the result produced by the numerical solver. First, if the result is SAT, it is possible that the satisfying assignment is not a true solution because of the approximations introduced by smoothing. Second, when the result is UNSAT, it is possible that the numerical algorithm ran into a local solution since gradient-based algorithms are incomplete in the presence of non-convex functions. 

To overcome these issues with the numerical solver, we have a  SAT solver component that gradually fixes values to the Boolean expressions in $\psi$. Fixing values to Boolean expressions will eliminate the approximations introduced by smoothing because, now, we do not have to smooth for any discontinuities these Boolean expressions may cause. Moreover, it  allows the numerical solver to focus the search in the region where the assignments to the Boolean expressions are satisfied.

 In order to do this, we had to make several changes to the SAT solver. A typical SAT solver interface takes in a problem, converts to CNF constraints, solves it and finally produces SAT/UNSAT. But, for the interaction with the numerical solver, the SAT solver needs to be incremental similar to the DPLL(T) solvers. However, there are two main differences between our approach and the DPLL(T) approach. First, in DPLL(T), the theory solver throws  conflicts whenever the current assignment to the theory atoms is unsatisfiable. However, in our case, because the numerical solver can run into local solutions, the conflicts are not strong.  Therefore, in \tech{}, the numerical solver throws \emph{soft conflicts} instead. Second,  when the current assignment is satisfiable, the theory solver in DPLL(T) approach also returns a list of consequences for any new variables that are forced by current assignment. In \tech{}, when a satisfiable solution is found by the numerical solver, this solution can be used to set values for the Boolean expressions and because of the closeness property of the numerical abstraction, most of these assignments can contribute to producing the final correct solution. However, it is not possible to mark these assignments as strong consequences because a different solution will produce different assignments to the Boolean expressions. Hence, in \tech{}, we treat these assignments as \emph{suggestions} that the SAT solver can use.   These suggestions are treated as new decisions in the SAT solver, and can be retracted if it resulted in a conflict later.

Thus, the SAT solver in \tech{} has two new data-structures.
\begin{itemize}
\item \textbf{Soft Learnts}. This data-structure is similar to the learnts data-structure used for normal conflicts. Soft learnts data-structure keeps the list of all learnt clauses arising from soft-conflicts as well as any other conflict that arises due to these soft learnts. 

\item \textbf{Suggestions}. Suggestions data-structure is an ordered list of assignments to the Boolean expressions suggested by the numerical solver. 
\end{itemize}

The SAT solver provides the following operations that the \tech{} core can call:
\begin{itemize}
	\item \textbf{Init($\psi_B$).} Initialize the problem with the constraints from $\psi_B$.
	\item \textbf{SolveIncremental().} Does incremental SAT solving by setting values to new variables and doing unit propagations. When picking a new variable to set, the SAT solver first tries the assignments in the suggestions data-structure. The method stops once an interface variable is set or if the problem is unsatisfiable. The output is the list of new interface mapping along with SAT / SOFT\_UNSAT / UNSAT. A SOFT\_UNSAT is an UNSAT  that is caused due to soft learnts. 
	\item \textbf{AddSoftConflict(conflict clause).} Performs conflict analysis and adds the learnt clause to the soft learnts data-structure. 
	\item \textbf{RemoveSoftLearnts().} Clears  soft learnts.
	\item \textbf{SetSuggestions(list of suggestions).} Populates the suggestions data-structure. 
	\item \textbf{RemoveSuggestions().} Clears suggestions.
	\item \textbf{Restart().} Backtracks all assignments made by the SAT solver so far.
\end{itemize}

With these components, we can, now, describe the actual interaction algorithm used by the \tech{} core.

\subsection{\tech{} core}\seclabel{core}

\newcommand{\sat}{\mathcal{S}}
\newcommand{\num}{\mathcal{N}}

\newcommand{\solve}{\textsc{Solve}}
\newcommand{\genSatSuggestions}{\textsc{GenSugesstions}}
\newcommand{\genUnsatSuggestions}{\textsc{GenUnsatSuggestions}}
\newcommand{\genConflict}{\textsc{GenConflict}}
\newcommand{\isConflict}{\textsc{IsConflict}}
\newcommand{\suggestions}{s}
\newcommand{\conflict}{c}
\begin{figure}[t]
	\begin{algorithmic}[1]
		\Statex \underline{$\solve$}($\psi$):
		\State $\psi_B = \LLB(\psi)$
		\State $\sat$.Init($\psi_B$)
		\State $\I = $ Empty
		\While {true}
			\State $\psi_N = \LLN(\psi, \I)$ \Comment{Phase 1}
			\State (res, $\ctrls_{\rhole}$) = $\num$.Solve($\psi_N^I$)
		\If {res = SAT} \Comment{Phase 2}
			\State $\suggestions$ = $\genSatSuggestions$($\ctrls_{\rhole}$)
			\State $\sat$.\t{SetSuggestions}($\suggestions$)
		\Else
			\State$\sat$.\t{RemoveSuggestions}( )
			\If {$\isConflict(\I)$}
				\State $\conflict$ = $\genConflict$( )
				\State$\sat$.\t{AddSoftConflict}($\conflict$)
			\EndIf
		\EndIf
		\State (res, $\I'$) =  $\sat$.\t{SolveIncremental}( ) \Comment{Phase 3}
		\If {res = SAT}
			\If {$\I= \I'$}
				\State \Return SAT
			\Else
				\State $\I \leftarrow \I'$
			\EndIf
		\EndIf
		\If {res = UNSAT}
			\State \Return UNSAT
		\EndIf
		\If {res = SOFT\_UNSAT} 
			\If {num\_restarts $<$ RESTART\_LIMIT}
				\State $\sat$.\t{RemoveSoftLearnts}( )
				\State $\sat$.\t{Restart}( )
				\State $\I$ = Empty
			\Else 
				 \State \Return UNSAT
			\EndIf
		\EndIf
		\EndWhile
	\end{algorithmic}
\caption{\tech{} algorithm.}
\figlabel{alg}
\end{figure}

The algorithm that \tech{} uses to  solve $\psi$ is shown in \figref{alg}. We use $\sat$ to denote the SAT solver and $\num$ to denote the numerical solver. The algorithm starts by initializing $\sat$ with the Boolean abstraction $\psi_B$. The interface mapping $\I$ is initialized to Empty meaning all Boolean expressions are assigned to $\bot$. Then, there are three phases that are performed repeatedly in a loop until a solution is found or the problem is detected to be unsatisfiable or the resource limits are reached. At a high-level, the first phase runs $\num$ on the numerical abstraction $\psi_N$ for the current $\I$, the second phase updates the SAT solver data-structures based on the result from $\num$, and the third phase does  more SAT solving to create a new $\I$. 

In the first phase, similar to~\cite{smooth}, \tech{} runs the numerical solver multiple times for various values of the smoothing parameter $\beta$. It starts with a small value for $\beta$ i.e. more smoothing and gradually increases the value of $\beta$ until a certain limit is reached, each time the numerical solver starting its search from the assignment found for the previous $\beta$ value. The numerical solver returns the solution found in the final iteration. 

In the second phase, if the result returned by $\num$ is SAT, the core uses the satisfying assignment to generate suggestions for the SAT solver. 
\figref{suggestionsAlg} shows the algorithm \tech{} uses to generate suggestions. The algorithm iterates over each Boolean expression in $\I$ whose value is not yet set and evaluates its numerical abstraction on the current assignment producing $d$ (Line \ref{lst:line:pdist}). Then, the algorithm suggests the value of this Boolean expression to be $d \ge 0$ because recall that $d$ actually represents the evaluation of a  smooth-approximation to the \dist{} function. For a Boolean unknown, the algorithm similarly evaluates its abstraction on the current assignment. However, in this case, $d$ is a value in $[0,1]$ and the algorithm suggests that the Boolean unknown is True if $d \ge 1/2$. Each suggestion is also associated with a cost and the SAT solver tries the suggestions in the increasing order of cost. Our algorithm gives lower cost to Boolean expressions/unknowns that are most uncertain (i.e. close to a branch) as they are more likely reasons for introducing inaccuracies in the numerical abstraction and hence, the SAT solver tries them first.

\begin{figure}[t]
	\begin{algorithmic}[1]
		\Statex \underline{$\genSatSuggestions$}($\ctrls_{\rhole}$, $\I$):
		\State $\suggestions$ = Priority List 
		\For {$\be \in \I $ such that $\I(\be) = \bot$}
		\State d = $\sem{\alpha_N(\be)}{\ctrls_{\rhole}}$ \label{lst:line:pdist}
		\State Add $(\be \leftarrow d \ge 0)$  to $\suggestions$ with cost = $|d|$
		\EndFor
		\For {$\bhole \in \I$ such that $\I(\bhole) = \bot$}
		\State d = $\sem{\alpha_N(\bhole)}{\ctrls_{\rhole}}$
		\State Add $(\bhole \leftarrow d \ge 1/2)$ to $\suggestions$ with cost = $|d - 1/2|$
		\EndFor
		\State \Return $\suggestions$
	\end{algorithmic}
	\caption{Algorithm to generate suggestions.}
	\figlabel{suggestionsAlg}
\end{figure}

If the result returned by $\num$ is UNSAT, then there are two possible options: 1. the SAT solver can continue to set values for more Boolean expressions assuming that the numerical solver might have run into a local solution or 2. the core can generate a soft conflict that would make the SAT solver come up with a different setting to $\I$, hence, temporarily directing the search to a different region. The heuristic $\isConflict$ decides whether to do option 1 or 2. \tech{} uses a simple heuristic that checks if the number of Boolean expressions that have values $\neq \bot$ is greater than a \emph{conflict threshold} ($\eta = 5$) to generate a soft conflict.


 In phase 3, $\sat.$SolveIncremental( ) is called which tries to set more Boolean variables. If the outcome of this call is SAT and there is no change to $\I$, then it means that a correct solution is found and the recent state of the SAT solver and the numerical solver will provide the values for $\ctrls_{\bhole}$ and $\ctrls_{\rhole}$. On the other hand, if the result is UNSAT (due to normal conflicts), then it means that $\psi_B$ is unsatisfiable which implies $\psi$ is also unsatisfiable. If the result is SOFT\_UNSAT, this means that the soft conflicts led the SAT solver to detect an unsatisfiability. Since the soft conflicts are just temporary conflicts, the algorithm clears the soft conflicts and restarts the SAT solver so that the SAT solver can make different decisions this time. When the number of restarts exceeds the RESTART\_LIMIT, a SOFT\_UNSAT is treated as an UNSAT. 
In all other cases, the algorithm continues the while loop with the new $\I$ generated by the SAT solver.

\subsection{Properties of \tech{}}
If \tech{} produces a solution $\sigma$ to $\psi$, then $\psi(\sigma)$ is true. Note,  only smoothing approaches such as~\cite{smooth} do not have this property because of the approximations introduced during smoothing.

However, there are situations where \tech{} may not find a solution even if there exists a solution. In practice, however, a right balance between the conflict threshold $\eta$ and the RESTART\_LIMIT allows \tech{} to solve some of the complex synthesis problems as shown in \secref{eval}.

%% file: interface.tex
\centering
\begin{tikzpicture}[node distance = 2cm, auto]

	\node[block, text width=4.6em, node distance=3.2cm](set){Incremental Solver};
	\node[block, right of = set, text width=2.8em, node distance=1.8cm](softconflicts){Soft Learnts};
	\node[block, below left of = softconflicts, text width=4.6em, node distance=1.5cm](suggestions){Suggestions};
	\node[outer,fit=(set) (softconflicts) (suggestions)] (sat) {};
	\node[block, left of=sat, text width = 5.2em, node distance =5cm, minimum height =1.5cm](encoder){\tech{} core\\ ($\I$)};
    \node [block, below of=encoder, text width=5.2em, node distance = 3cm, minimum height=1.5cm] (num) {Numerical solver};
    \node [coordinate, right of=num] (output) {};
    \node (text) [anchor=south] at ([yshift=-2.0em]sat.south) {SAT solver ($\psi_B$)};
     
	 \path[arrow] ([yshift=4ex]sat.west) edge node[above]{ $\I$ } node [below]{$\ctrls_{\bhole}$} ([yshift=4ex]encoder.east);
	 \path[arrow] ([yshift=-3ex]encoder.east) edge node[above]{soft conflicts} ([yshift=-3ex]sat.west);
	 \path[arrow] ([xshift=4ex]encoder.south) edge node[right]{ $\psi_N^{\I}$ }  ([xshift=4ex]num.north);
	 \path[arrow] ([xshift=-3ex]num.north) edge node[left, rotate=90, yshift=2ex, xshift=5ex]{(UN)SAT} node[right]{$\ctrls_{\rhole}$} ([xshift=-3ex]encoder.south);
	  \path[arrow] ([yshift=-5ex]encoder.east) edge [out=-30,in=-150] node[below] { suggestions } ([yshift=-5ex]sat.west);
    
\end{tikzpicture}
\caption{Overview of the solver interface.}
\figlabel{overview} 

%% file: eval.tex
\section{Evaluation}\seclabel{eval}
In this section, we evaluate \tech{} on 5 case-studies involving hybrid systems. In particular, we focus on answering the following questions: (1) Can \tech{} solve complex synthesis problems that involve a combination of discrete and continuous reasoning? (2) How does it compare with only smoothing techniques?  (3) How does it compare with existing SMT solvers and mixed integer approaches?

\paragraph{Experimental Setup}
\tech{} uses the SNOPT software~\cite{gill2005snopt} for performing numerical optimization. SNOPT uses sequential quadratic programming and can handle constrained optimization better than standard gradient descent techniques.  For the SAT solver, we took the MiniSAT solver and modified it  as described in \secref{solver}.
All experiments are run on a machine using 2.4GHz Intel i5 core with 8GB RAM. 
\begin{figure*}[ht]
	\begin{tabular}{|l| c c c c c| c c c| c c | c c c|}
		\hline
		{} & \multicolumn{5}{c|}{Stats} & \multicolumn{5}{c|}{\tech{}} & \multicolumn{3}{c|}{only smoothing} \\
		\hline
		{Benchmarks} & {\#$\bhole$} & {\#$\rhole$}	&{\#I} & {\#B} & {\#A} & \multicolumn{3}{c|}{Time(s)} & {\#N} & {\#R} & {\#S} & {\#C} & {$T_C$}\\ 
		\cline{7-9}
		{} & {} & {} & {} & {} & {} & {20\%} & {median} & {80\%} & {} & {} & {} & {} & {}\\
		\hline
		{LaneChange} & {12} & {22} &{50}& {11455} & {2350} & {276s} & {634s} & {1261s} & {94} & {4} & {9/300} & {0/300} & {> 2500s}  \\
		\hline
		{QuadObstacle} & {4} & {38} & {70} & {1753} & {1714} & {81s} & {217s} & {822s} & {16} & {1} & {34/300} & {21/300} & {857s} \\
		\hline
		{QuadLanding} & {6} & {44} & {60} & {1567} & {1413} & {231s} & {374s} & {637s} & {32} & {2} & {62/300} & {22/300} & {665s} \\
		\hline
		{ParallelPark} & {9} & {15} & {100} & {11204} & {1904} & {354s} & {630s} &{870s} & {98} & {5} & {0/300} & {0/300} & {> 2100s} \\
		\hline
		{Thermostat} & {0} & {2} & {500} &  {30908} & {2100} & {78s} & {175s} & {280s} & {31} & {3} & {21/300} & {19/300} & {220s} \\
		\hline
	\end{tabular}
	\caption{Benchmarks statistics and results. \#$\bhole$: number of Boolean holes, \#$\rhole$: number of real holes, \#I: number of iterations used in the simulation, \#B: number of Boolean expressions, \#A: number of assertions, \#N: number of calls to numerical solver, \#R: number of restarts of SAT solver, \#S: number of times a solution is found, \#C: number of times a correct solution is found and $T_C$ is the expected time for the only smoothing approach to find a solution with 90\% confidence. }
	\figlabel{results}
\end{figure*}

\subsection{Benchmarks}
 Apart from the lane changing example in \secref{overview}, we used \tech{} to synthesize two benchmarks involving a 1-dimensional quad-copter, the parallel parking benchmark from~\cite{chaudhuri2012euler}, and the thermostat benchmark from~\cite{jha2010synthesizing}. \figref{results}  lists the 5 benchmarks together with some statistics such as the number of Boolean and real unknowns, the number of iterations the controller (to be synthesized) is simulated in the specification, and the total number of Boolean expressions and assertions. Full benchmark problems along with the demos of the synthesized solutions can be found in the supplementary material.
 \paragraph{Quadcopter obstacle  avoidance} The task is to synthesize a controller to perform the maneuver shown in \figref{quad} without colliding with the obstacle. The program with unknowns for this controller is shown in \figref{quad_code}. This controller has three modes where each mode uses a proportional-derivative (PD) controller to set the forces that need to be generated by the two rotors of a simplified 1-D quadcopter.  The synthesizer is required to find the switching conditions (which are based on the position of the copter similar to the lane change benchmark) as well as the parameters of the PD controllers for the different modes. Note that, in this case, a single PD controller is not sufficient to perform the task and hence, it is necessary to compose multiple PD controllers as shown in the template. The synthesizer is also required to find the values of  the intermediate desired states for the PD controllers.

 \begin{figure} 
	\includegraphics[width=0.40\textwidth]{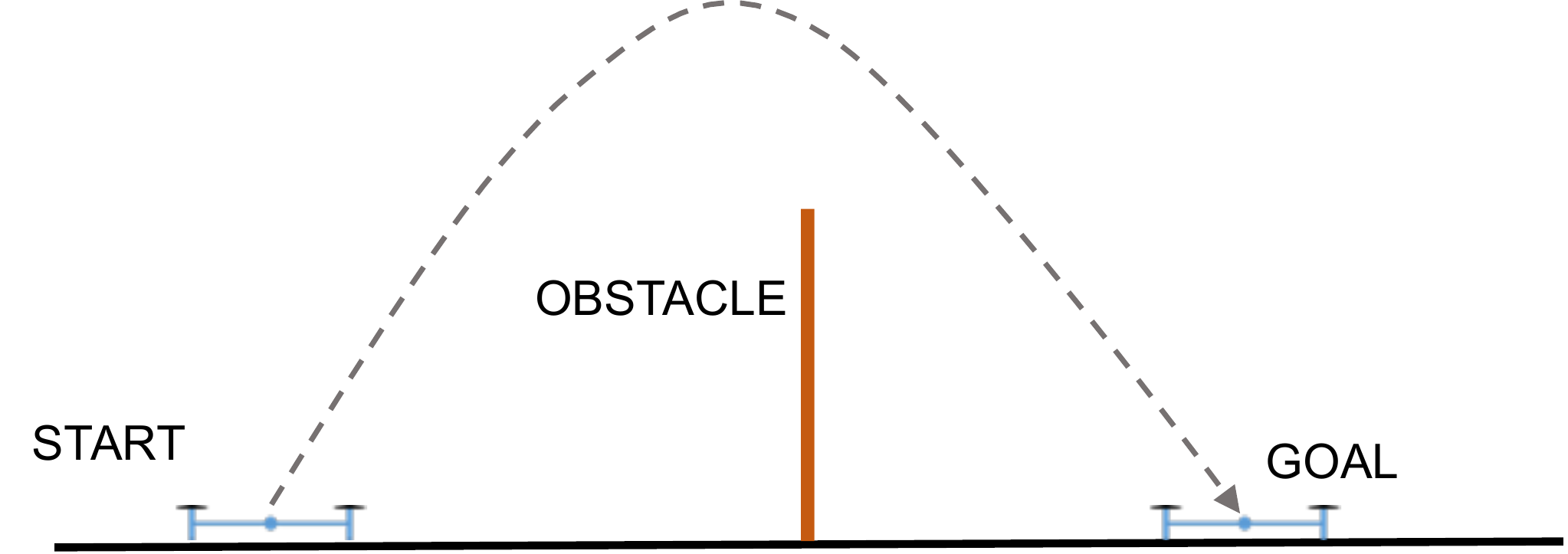}
	\caption{Quadcopter obstacle avoidance scenario.}
	\figlabel{quad} 	
 \end{figure}

 \begin{figure}
 \begin{lstlisting}
 void Controller(Copter c) {
 	if (genSwitchExp(c)) { /* mode 1 */
 			genPDController(c);
 	} else if (genSwitchExp(c)) { /* mode 2 */
 			genPDController(c);
 	} else { /* mode 3 */
 			genPDController(c);
 	}
 }
 
 generator void genPDController(Copter c) {
 	F = (c.y - ??%*$_r$*))*??%*$_r$*) + (c.vel.y - ??%*$_r$*))*??%*$_r$*);
 	bias = (c.ang - ??%*$_r$*))*??%*$_r$*) + (c.angvel - ??%*$_r$*))*??%*$_r$*) 
 		     + (c.x - ??%*$_r$*))*??%*$_r$*) + (c.vel.x - ??%*$_r$*))*??%*$_r$*);
 	c.left_force = F + bias;
 	c.right_force = F;
  }
 \end{lstlisting}
 \caption{Program with unknowns for quadcopter controller.}
 \figlabel{quad_code}
 \end{figure}
 
 \paragraph{Quadcopter landing} Using  the same template shown in \figref{quad_code}, but with  different specifications, it is possible to synthesize controllers for achieving other goals. In this benchmark, we synthesize a controller for landing a quadcopter gracefully. The copter starts at a position above the ground with an initial thrust that imbalances the copter and target is to reach a position on the ground without crashing.  There are no obstacles (other than the ground) in this case, but the synthesizer still needs to figure out how to  compose the different PD controllers to achieve the goal.

 \paragraph{Parallel parking} This benchmark synthesizes a controller to parallel park a car as described in~\cite{chaudhuri2012euler} (a tool based on the smooth interpretation work~\cite{smooth}). The template for this benchmark is similar to the template for the lane changing benchmark. Our template is different from~\cite{chaudhuri2012euler} in two aspects. \cite{chaudhuri2012euler} uses switching conditions based on time; we replaced them with conditions based on the state of the car since it leads to more robust controllers.~\cite{chaudhuri2012euler} only uses 10 time-steps to do the simulation, but in our template, we decrease the step size and increased the number of simulation steps to 100. This reinforces the fact that our technique scales much better than the smoothing technique that~\cite{chaudhuri2012euler} uses.

 \paragraph{Thermostat} The final benchmark is to synthesize the thermostat controller described in~\cite{jha2010synthesizing}. This thermostat is a state machine with four states: OFF, HEATING, ON and COOLING. The switching conditions for transitioning from HEATING to ON and COOLING to OFF are fixed by the constraints of the thermostat's heater. The other two switching conditions should  be figured out by the synthesizer  such that the temperature of the room is maintained between 18$^\circ$C and 20$^\circ$C. In addition to the safety conditions, this benchmark encodes some performance metrics, in particular, it adds minimum dwell time constraints for the OFF and ON phases. The minimum dwell time constraint states that the thermostat should at least be in the state for $T$ seconds before transitioning to the next state.  The system in \cite{jha2010synthesizing} takes in these universal constraints (that should be true in every state) and uses a fix-point computation based algorithm to find the switching conditions. In \tech{}, we instead specify the problem by simulating the thermostat for 500 time steps with dt = 2s  and asserting that the constraints are satisfied at every time step. Most of the SMT solvers  choke when given a problem of this magnitude, but \tech{} is still able to synthesize it. \figref{thermostat} shows how the room temperature and the state of the thermostat change with a controller that is synthesized with minimum dwell time constraint of 200s for the OFF and ON phases. 
 
 \begin{figure}
 	\includegraphics[width=0.48\textwidth]{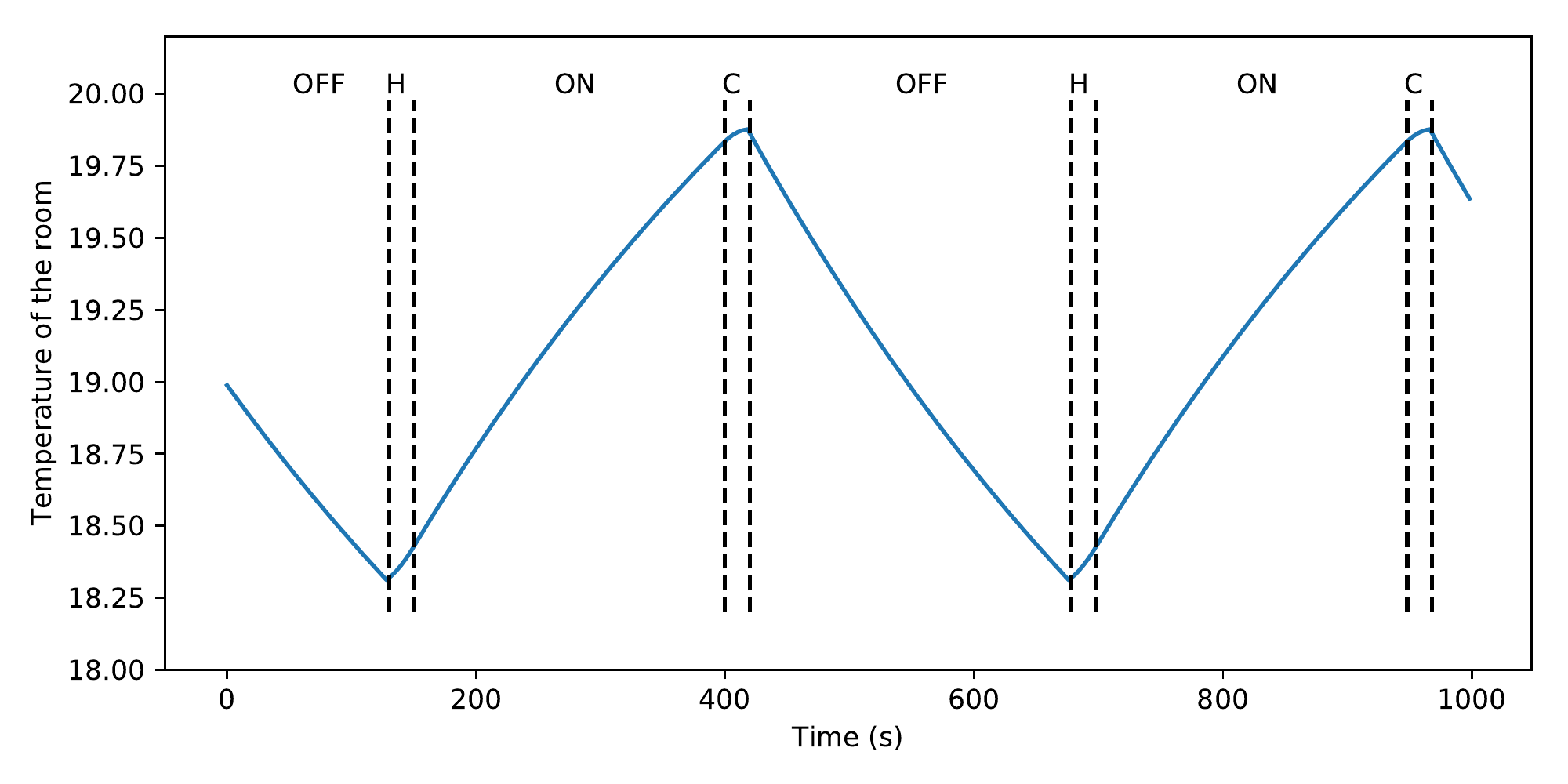}
 	\caption{Room temperature vs time for the synthesized thermostat controller. }
 	\figlabel{thermostat}
 \end{figure}
 
 \subsection{Results}
\figref{results} shows the evaluation results. The \C{Time} column lists the  time taken in seconds (20th percentile,  median, and 80th percentile) to synthesize the benchmarks in \tech{}. This experiment is run with a conflict threshold $\eta = 5$ and unlimited RESTART\_LIMIT, but with a timeout of 30 minutes. We ran each benchmark 10 times and \tech{} is able to synthesize all the benchmarks for all 10 runs within 25 minutes. The \C{#N} and \C{#R} columns show the median number of numerical iterations and SAT solver restarts taken by these benchmarks. 

Next, we performed an experiment to compare \tech{} with an only smoothing approach. For each benchmark, we took the numerical abstraction when interface mapping $\I$ is empty and ran our numerical solver on this abstraction with a random initial assignment to the unknowns. We ran each benchmark 300 times and collected the number of times the numerical solver is able to converge to a valid solution for the abstraction (shown in \C{#S} column in \figref{results}). Since, some of the solutions might actually be incorrect on the original problem due to the approximations introduced by smoothing, we also computed the number of times a correct solution is found by the only smoothing approach (shown in \C{#C} column). Using these statistics and the average time to run each numerical iteration, we also computed the expected time for the only smoothing approach to find a correct solution with 90\% confidence (shown in $T_C$ column). The only smoothing approach can synthesize the quadcopter and thermostat benchmarks if the numerical solver is run sufficient number of times, but it is not able to find a correct solution for the lane change and parallel parking benchmarks even with 300 iterations. This shows that searching the numerical space based on the program structure is very important for some of these benchmarks. Note that, in this experiment the comparison is done against the smoothing approach described in this paper and not the smooth interpretation work in~\cite{smooth}. 

We also performed an experiment to see how the conflict threshold $\eta$ effects the performance of \tech{}.  A higher conflict threshold means that  the search space of the numerical solver is significantly reduced by the assignments to a higher number of Boolean expressions, which increases the likelihood for the numerical solver to hit the global solution (if there is one in the reduced search space). On the other hand, a higher conflict threshold will  result in spending more time to eliminate a truly infeasible partial assignment because the system needs to wait until the conflict threshold is hit. \figref{conflict} shows the graph for the time as well as the number of numerical iterations required for synthesizing the benchmarks when $\eta = [3, 4, 5, 6, 7]$. It can be seen that different benchmarks have different behaviors since the trade-off between the two aspects mentioned above depends on the structure of the benchmark. However, we can see that for any of the small conflict thresholds ($3 \le \eta \le 7$), the system can synthesize the benchmarks in reasonable time.  

\begin{figure}
	\includegraphics[width=0.23\textwidth]{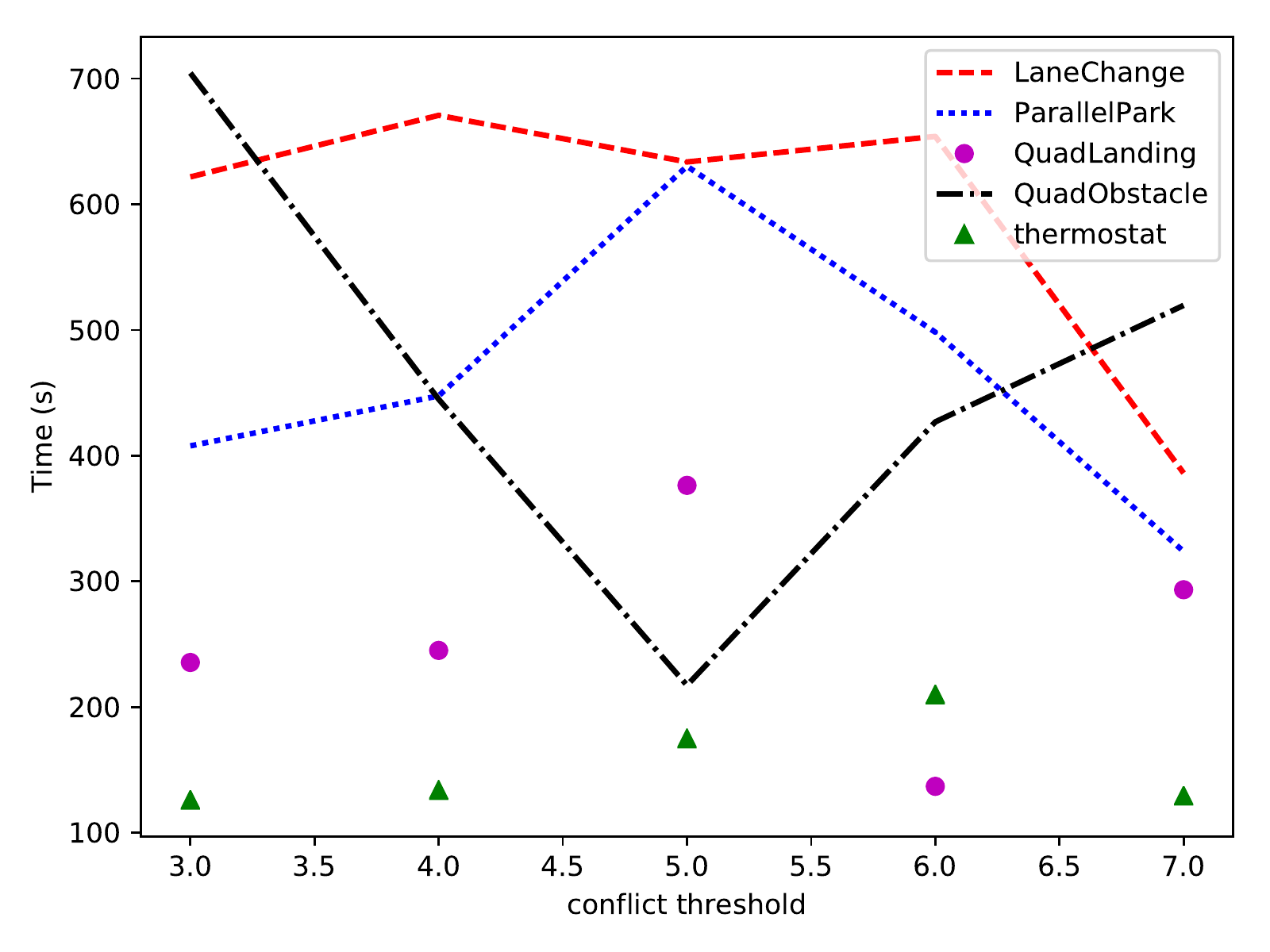}
	\includegraphics[width=0.23\textwidth]{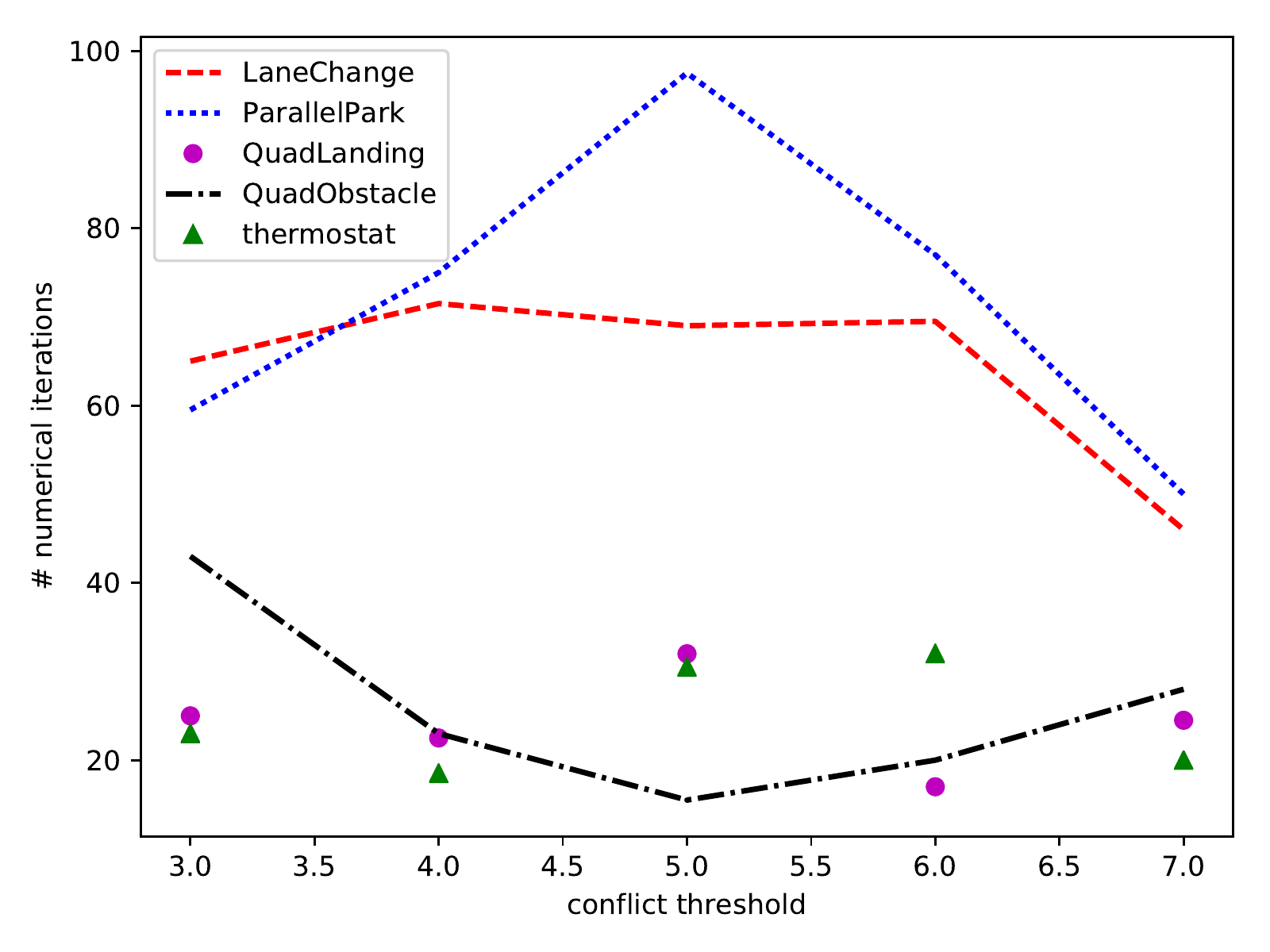}
	\caption{Graphs showing the time and the number of numerical iterations vs the conflict threshold in \tech{}.}
	\figlabel{conflict}
\end{figure}

\subsection{Comparison to SMT solvers}
We compared \tech{} with dReal, a state-of-the-art SMT solver for reals. In-order to do this comparison, we wrote a script to translate the benchmarks written in \tech{}  into the SMT-lib format and fed them to the dReal solver. However, dReal could not solve any of our benchmarks with a timeout of 60 minutes. We also ran Z3 on the thermostat benchmark (the only benchmark in our suite that does not contain non-linear functions) and found that even Z3 could not solve with a timeout of 60 minutes. These results support our claim that the traditional SMT approaches do not scale for these kinds of synthesis problems.  

\subsection{Comparison to mixed integer approaches}
In the final experiment, we evaluate how the mixed integer approach compares to \tech{}. Since most of the mixed integer solvers only handle linear and convex constraints, we created a toy version of the lane change benchmark.  We replaced the complex dynamics of the car with a simple point car model that can only either move along the x-direction or the y-direction. To deal with conditionals and disjunctions, we manually translated them to linear constraints using the big-M method. This translation, however, introduces one 0-1 integer variable for each Boolean expression in the original problem and a continuous variable for every conditional. We  verified that the translation to the mixed integer format is correct by fixing the values to the unknowns with a solution found by \tech{}. 
We ran this translated version using Gurobi, a state of the art mixed integer solver. However, Gurobi was not able to find a solution with a timeout of 60 minutes, whereas \tech{} was able to solve the benchmark in just 3 seconds.  This shows that even the mixed integer solvers are not suitable for handling these benchmarks. 

%% file: related.tex
\section{Related Work}
To the best of our knowledge, this paper is the first that achieves both  end-to-end differentiability of a program and  uses a SAT solver to help perform numerical optimization to find unknowns in a program. However, there are several related works for each of the two  pieces. 

\noindent
\textbf{End-to-end differentiability:} The idea of achieving end-to-end differentiability in not new in the neural networks community. The entire back-propagation algorithm is  based on this principle. Our idea of smoothing using sigmoids is inspired by the Neural networks' use  of sigmoids in-place of step functions to make the network differentiable. For neural networks, there are libraries such as TensorFlow~\cite{45381} where users can write their  networks in a high-level language and the library can automatically compute the required gradients. In \tech{}, we use similar ideas to programmatically smooth conditionals and conjunctions. The recent works on neural Turing machines~\cite{graves2014neural} and neural program interpreters~\cite{reed2015neural, kurach2015neural, neelakantan2015neural, kaiser2015neural}  achieve end-to-end differentiability  in the presence of discrete structure. They accomplish this by turning the discrete variables into continuous variables for encoding the probabilities of the different discrete options. On the other hand, in this paper, we relax Boolean unknowns to  real unknowns in the range [0, 1] and then, use a SAT solver to fix their values. This approach allows us to benefit from the fact that SAT solvers are inherently better at handling discrete problems.

Smoothing in the context of programs is introduced in the smooth interpretation work~\cite{smooth} and subsequently used in~\cite{chaudhuri2014bridging} to solve synthesis problems involving Boolean and quantitative objectives. Smoothing using sigmoids that we use in this paper is a simplified version of the smoothing algorithm used in~\cite{smooth}. However, our approach allows us to use automatic differentiation techniques to compute gradients necessary for the numerical optimization.

\noindent
\textbf{SAT/SMT solvers:} There have been many approaches to SMT solving over real numbers that
incorporate numerical methods. Examples include convex optimization
algorithms~\cite{BorrallerasLNRR09,DBLP:conf/fmcad/NuzzoPSS10,smc},
interval-based algorithms~\cite{HySAT,DBLP:conf/cade/GaoKC13},
Bernstein polynomials~\cite{bern}, and linearization
algorithms~\cite{DBLP:conf/cade/CimattiGIRS17}. However, these approaches  strictly partition the problem into a Boolean part and  many of numerical parts that loses the structural dependencies between the numerical parts and hence, they are not able to leverage the benefits of doing numerical optimization on the entire problem. 

\tech{}'s modifications to the SAT solver to support \emph{soft learnts} and \emph{suggestions} is similar to the idea of assumptions in~\cite{nadel2012efficient, audemard2013improving}. These works use assumptions to support incremental SAT solving when there are constraints that only hold for one invocation. In our approach, we use soft learnts and suggestions  to inform the SAT solver that they can be revoked any time if a conflict is detected. 

\noindent
\textbf{Hybrid Systems:} The hybrid systems community have also looked at problems involving discrete and continuous components. Some of the recent works in this area include~\cite{prabhakar2017formal,ozay2013computing,girard2012controller}. These approaches use a discrete abstraction of the continuous components and  perform purely discrete reasoning. On the other hand, our approach shows that by leveraging the numerical structure of the problem, it is possible to scale hybrid systems synthesis to very complex problems.  

\noindent
\textbf{Control Optimization:}
The control optimization community usually uses mixed integer programming to solve these kinds of problems. For example,~\cite{deits2015efficient} uses the mixed integer approach for motion planning in UAVs. Similarly, mixed integer programming is used extensively in model predictive control~\cite{camacho2013model}. In these approaches the task is to directly learn the actions for every time step rather than a program for the controller. Learning a program introduces more discreteness and as shown in the evaluation, mixed integer approaches do not work well for the benchmarks in this paper.